%% file: 00-paper-mldb_template.tex
\documentclass[preprints,journal,accept,pdftex,moreauthors]{Definitions/mdpi} 
\usepackage{booktabs}
\usepackage{longtable}
\usepackage{array}
\usepackage[table]{xcolor}
\usetikzlibrary{arrows.meta,calc,positioning}

\firstpage{1} 
\pubvolume{1}
\issuenum{1}
\articlenumber{0}
\pubyear{2026}
\copyrightyear{2026}
\datereceived{ } 
\daterevised{ } % Comment out if no revised date
\dateaccepted{ } 
\datepublished{ } 
\Title{Cross-Scale Spatial Reasoning: A Learning-Centered Framework for Designing Immersive Learning Environments}

\Author{
Dishita Turakhia $^{1}$\orcidB{}
}

\AuthorNames{Dishita Turakhia}

\address{%
$^{1}$ \quad Department of Computer Science, Emory University; dishita.turakhia@emory.edu
}

\corres{Correspondence: dishita.turakhia@emory.edu}

\abstract{
Understanding complex scientific and engineering systems often requires learners to connect structures, processes, and behaviors across different spatial scales and levels of organization. Although research on spatial cognition, scale, multiple representations, and systems thinking has examined aspects of this challenge, these perspectives do not always provide explicit guidance for designing learning activities that require learners to integrate information across levels. Immersive technologies offer opportunities to explore otherwise inaccessible structures, coordinate representations, and investigate system behavior, but their educational value depends on the reasoning they support. This paper proposes Cross-Scale Spatial Reasoning (CSSR) as an integrative framework for understanding how learners establish and use spatial and structural relationships across scales or organizational levels. We describe five interrelated processes: scale navigation, cross-scale relational mapping, representational coordination, dynamic and causal reasoning, and consequence-oriented reasoning. Building on these processes, we develop a learning-centered framework that connects reasoning objectives to immersive design mechanisms, instructional scaffolding, and assessment. Illustrative applications in biology, semiconductor design, manufacturing, and architecture show how the framework can inform domain-specific learning activities. We also propose testable research propositions and approaches for evaluating cross-scale understanding, instructional effectiveness, and transfer. Rather than treating immersion as an educational outcome in itself, CSSR provides a basis for designing and studying learning environments that help learners explain how complex systems work across levels.
}


\begin{document}

%%%%%%%%%%%%%%%%%%%%%%%%%%%%%%%%%%%%%%%%%%
% \setcounter{section}{-1} %% Remove this when starting to work on the template.
% \section{How to Use this Template}

% The template details the sections that can be used in a manuscript. Note that the order and names of article sections may differ from the requirements of the journal (e.g., the positioning of the Materials and Methods section). Please check the instructions on the authors' page of the journal to verify the correct order and names. For any questions, please contact the editorial office of the journal or support@mdpi.com. For LaTeX-related questions please contact latex@mdpi.com.%\endnote{This is an endnote.} % To use endnotes, please un-comment \printendnotes below (before References). Only journal Laws uses \footnote.

% % The order of the section titles is different for some journals. Please refer to the "Instructions for Authors” on the journal homepage.

\input{sections/01-introduction}

\input{sections/02-Theory}

\input{sections/03-Concept}

\input{sections/04-Framework}

\input{sections/05-Applications}

\input{sections/06-Evaluation}

\input{sections/07-Discussion}
\input{sections/08-conclusion}

\vspace{6pt}

\reftitle{References}

% Please provide the correct journal abbreviation (e.g. according to the “List of Title Word Abbreviations” http://www.issn.org/services/online-services/access-to-the-ltwa/).

% \bibliography{your_external_BibTeX_file}
\bibliography{xx-References}
\end{document}

%% file: sections/01-Introduction.tex
\section{Introduction}
\label{sec:introduction}

Understanding complex scientific and engineering systems requires more than knowledge of their individual components. Learners must also explain how structures and processes interact across spatial scales and levels of organization to produce system behavior. In biology, molecular interactions contribute to cellular function; in manufacturing, local tool--material interactions influence the properties of finished components; in semiconductor design, device characteristics and physical interconnections affect circuit performance; and in architecture, structural connections influence the distribution of forces throughout a building. Despite differences in their underlying principles, these domains share a fundamental learning challenge: understanding how relationships across levels shape the behavior of a larger system.

This challenge is compounded by the ways complex systems are represented. Scientific and engineering concepts are often communicated through multiple forms of information, including three-dimensional structures, diagrams, physical models, and simulations. Each representation reveals particular properties or relationships, yet understanding one representation does not necessarily translate into understanding another. Moreover, moving between representations is not always equivalent to moving between spatial scales or organizational levels. Learners must recognize these distinctions, identify relevant relationships, and integrate information across views to develop coherent explanations of system behavior.

Research in spatial cognition, science education, and the learning sciences provides important foundations for understanding these demands. Spatial thinking supports the interpretation and transformation of spatial information across STEM disciplines~\citep{nrc2006spatial,uttal2012spatial}. Research on size-and-scale cognition examines learners' understanding of magnitude, proportion, and relationships among entities at different scales~\citep{tretter2006boundaries,cheek2017scale,magana2012fs2c}. Studies of complex systems have highlighted difficulties in connecting component-level interactions with system-level behavior~\citep{wilensky1999levels,jacobson2006complex,hmelosilver2004sbf}. Research on multiple representations further demonstrates the importance of coordinating complementary views and establishing meaningful correspondences among them~\citep{ainsworth2006deft,rau2017representations}.

Although these perspectives address important aspects of learning about complex systems, a central instructional challenge remains: \textbf{how can learning environments help learners establish and use relationships across scales and organizational levels, rather than understand each level in isolation?} A learner may identify the structures within a biological system without explaining how molecular interactions contribute to cellular behavior. Similarly, a student may understand the components and operations of a manufacturing process but struggle to connect local material interactions to the properties of the finished component. In both cases, the difficulty lies not necessarily in understanding individual structures or processes, but in integrating knowledge across levels to explain how the system works.

Immersive technologies, including augmented reality (AR) and virtual reality (VR), offer opportunities to support this integration. Interactive three-dimensional environments can make otherwise inaccessible structures available for exploration, connect complementary representations, and allow learners to investigate how systems respond to changes. Existing immersive learning systems demonstrate applications in biology, manufacturing, semiconductor fabrication, electrical engineering, and structural analysis~\citep{laubscher2024cell,conesa2024lathe,wang2024cleanroom,wolbach2024circuits,turkan2017structural}. However, making information visible and interactive does not guarantee that learners understand the relationships it represents. Research on immersive learning has reported varied outcomes, including cases where greater immersion increased presence without improving learning~\citep{makransky2019presence,parong2018science,coban2022meta}. These findings highlight the need to design immersive experiences around explicit learning objectives and the reasoning required to achieve them.

To address this challenge, we propose \textbf{\textit{Cross-Scale Spatial Reasoning}} (CSSR) as an integrative framework for understanding how learners establish and use spatial and structural relationships across physical scales or organizational levels. CSSR focuses on \textit{cross-scale integration}: connecting information across levels to explain system behavior and predict the consequences of changes. Rather than introducing entirely new cognitive abilities, the framework brings together established perspectives on spatial reasoning, relational mapping, representational coordination, and systems understanding to characterize the learning demands involved in constructing cross-scale explanations.

We identify five interrelated processes within CSSR: \textit{scale navigation}, \textit{cross-scale relational mapping}, \textit{representational coordination}, \textit{dynamic and causal reasoning}, and \textit{consequence-oriented reasoning}. Together, these processes describe how learners identify relevant levels, establish relationships among structures, coordinate information across representations, explain system behavior, and evaluate possible outcomes. The framework connects these reasoning demands to instructional design by identifying the immersive interactions, scaffolding strategies, and assessment approaches that may support cross-scale understanding.

\begin{figure}[t]
    \centering
    \includegraphics[width=\linewidth]{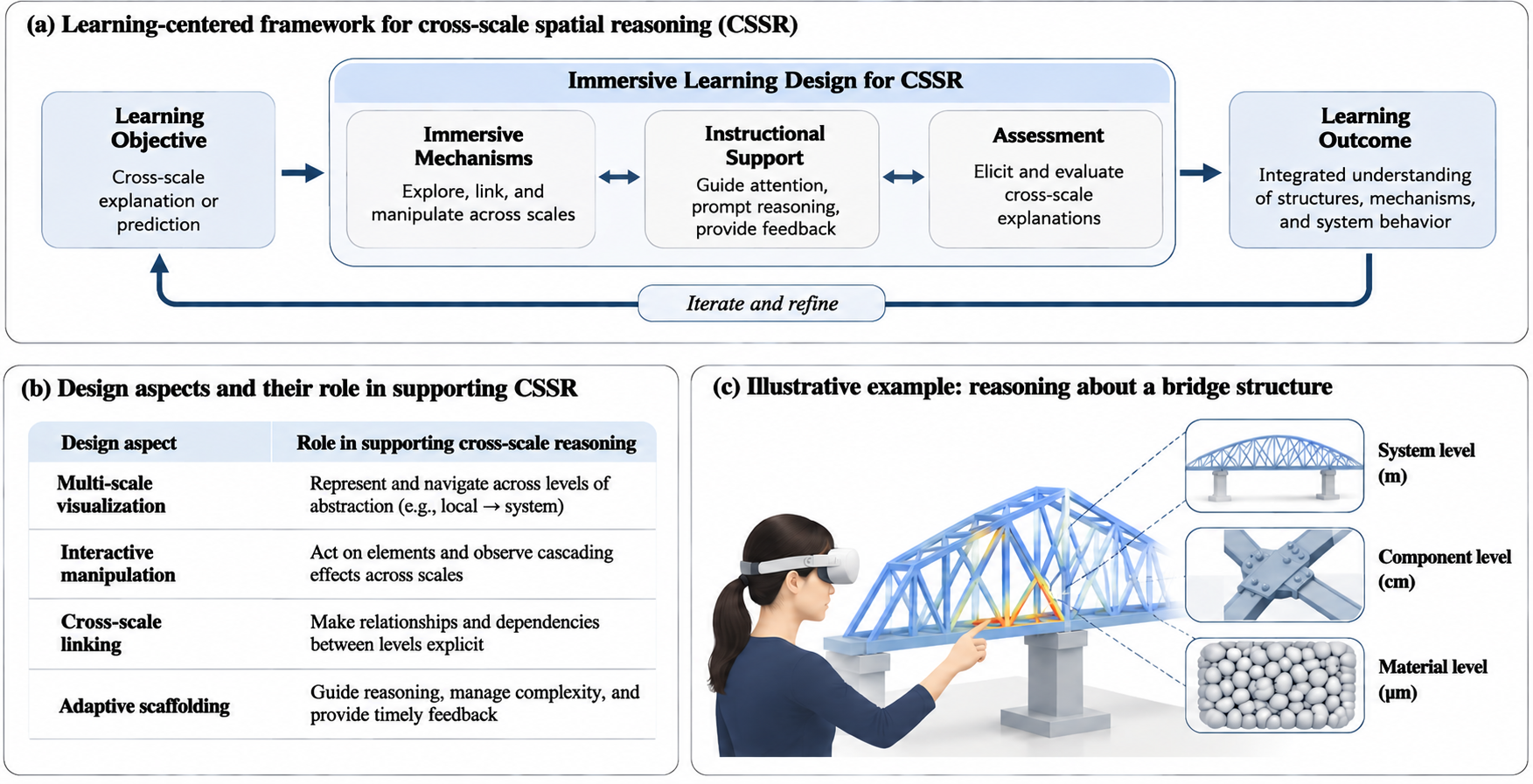}
    \caption{Overview of the proposed learning-centered framework for Cross-Scale Spatial Reasoning (CSSR). (a) The framework connects cross-scale learning objectives with immersive interaction mechanisms, instructional support, and assessment to promote integrated understanding of complex systems. (b) Key design mechanisms support exploration across scales, interactive manipulation, explicit connections between representations, and instructional scaffolding. (c) An illustrative immersive learning scenario in structural engineering shows how learners can investigate relationships among bridge-level behavior, structural connections, and material properties. The illustration represents a proposed educational scenario rather than an evaluated system.}
    \label{fig:cssr-overview}
\end{figure}

Figure~\ref{fig:cssr-overview} provides an overview of this learning-centered approach. It illustrates how cross-scale learning objectives can inform immersive design mechanisms, instructional support, and assessment, with an example from structural engineering demonstrating how learners might investigate relationships across system, component, and material levels.

\textbf{This paper makes three contributions.} First, it develops a conceptual account of CSSR by synthesizing established research on spatial cognition, scale, multiple representations, and complex systems, identifying five constituent processes and their conceptual boundaries. Second, it proposes a learning-centered framework that connects cross-scale reasoning objectives to immersive design mechanisms, instructional scaffolding, and testable research propositions. Third, it outlines approaches for assessing cross-scale understanding, evaluating instructional effectiveness, and investigating whether learners can apply this understanding to unfamiliar problems. Illustrative applications in biology, manufacturing, semiconductor design, and architecture demonstrate how these contributions can inform learning activities across different scientific and engineering contexts.

%% file: sections/02-Theory.tex
\section{Theoretical Foundations}
\label{sec:theoretical-foundations}

Understanding relationships across spatial scales and organizational levels draws on several established areas of research. Spatial cognition addresses how people interpret and transform spatial information, while size-and-scale cognition concerns magnitude, proportion, and relationships among entities of different sizes. Research on multiple representations examines how learners coordinate different forms of information, and systems thinking addresses how components and interactions contribute to larger-scale behavior. Research on immersive learning further examines how interactive environments can support these forms of understanding. This section brings these perspectives together to establish the theoretical foundations of CSSR.

\subsection{Spatial Cognition and Reasoning}
\label{sec:spatial-cognition}

Spatial cognition concerns how people perceive, represent, and reason about objects and their relationships in space. It encompasses abilities such as spatial visualization, mental rotation, and orientation, which are important for interpreting three-dimensional structures and understanding spatial configurations. The National Research Council characterizes spatial thinking through the integration of spatial concepts, representational tools, and reasoning processes~\citep{nrc2006spatial}. These abilities play an important role in STEM learning, particularly when learners must understand structures that are difficult to observe or manipulate directly~\citep{uttal2012spatial}.
Research has also demonstrated that spatial abilities can develop through targeted instruction and experience~\citep{newcombe2010early,uttal2013malleability}. External representations, including physical models, scientific diagrams, and interactive three-dimensional environments, can support spatial reasoning by making complex structures and relationships available for inspection. Such representations may reduce the need to reconstruct spatial information entirely through mental imagery, allowing learners to focus on relationships relevant to the task.
However, visualizing a structure and understanding its role within a larger system are different learning demands. A learner may accurately identify the geometry and arrangement of individual components without explaining how their interactions contribute to system behavior. Spatial cognition therefore provides an important foundation for cross-scale understanding, particularly when learners must relate spatial information across multiple levels.

\subsection{Reasoning Across Scales and Organizational Levels}
\label{sec:scale-reasoning}

Reasoning across scales requires distinguishing among physical magnitude, organizational level, and conceptual abstraction. Although these dimensions often overlap, they describe different relationships. \textit{Physical scale} concerns the size or spatial extent of an entity or phenomenon. \textit{Organizational level} concerns how components are organized into larger systems and how their interactions contribute to collective behavior. \textit{Conceptual abstraction} concerns the level of detail or the form in which information is represented. Moving between organizational levels may involve a change in physical scale, but moving between representations does not necessarily involve either.
Research on size-and-scale cognition has examined how learners understand magnitudes, proportional relationships, and phenomena beyond everyday experience. Tretter and colleagues~\citep{tretter2006boundaries} identified differences in students' and experts' understanding of scientific scales, emphasizing the importance of reference points and relative-size relationships. Jones and colleagues~\citep{jones2008scale} similarly investigated teachers' understanding of spatial scale, while Cheek, LaDue, and Shipley~\citep{cheek2017scale} examined the cognitive and educational challenges associated with spatial and temporal scale.
These findings have informed frameworks for developing and assessing scale understanding. Magana and colleagues~\citep{magana2012fs2c} proposed an integrated framework for size-and-scale cognition that addresses qualitative and quantitative relationships, including ordering, categorization, and proportional reasoning. More recently, Harper-Gampp and colleagues~\citep{harpergampp2026assc} developed and validated the Assessment of Size and Scale Cognition (ASSC), providing a foundation for evaluating learners' understanding of size and scale.

Reasoning across organizational levels introduces additional challenges. Wilensky and Resnick~\citep{wilensky1999levels} demonstrated that learners may confuse the behavior of individual components with the behavior of the larger systems in which they participate. For example, understanding how an individual vehicle moves does not, by itself, explain how traffic congestion develops across a road network. Explaining the larger phenomenon requires identifying the interactions among vehicles and how those interactions produce collective patterns.
The distinction between physical scale and organizational level is particularly important in scientific and engineering education. Explaining how molecular interactions contribute to cellular function involves changes in both physical scale and system organization. Understanding how structural elements contribute to the behavior of a building involves relationships among components and assemblies. By contrast, moving between a three-dimensional model and a schematic may change the representation without changing the underlying level of organization.

These cases illustrate that learning across scales is not simply a matter of changing magnification or moving between views. Learners must recognize which levels are relevant, understand their relationships, and determine how information at one level contributes to an explanation at another.

\subsection{Learning Across Multiple Representations}
\label{sec:multiple-representations}

Scientific and engineering systems are often described through multiple representations, each emphasizing different properties or relationships. Three-dimensional models communicate spatial structure, diagrams describe structural or functional connections, equations express quantitative relationships, and simulations illustrate changes over time. Understanding a complex system may require learners to interpret these representations individually and coordinate the information they provide.
Ainsworth's Design, Functions, Tasks (DeFT) framework explains how multiple external representations can complement one another, constrain interpretation, and support deeper understanding~\citep{ainsworth2006deft}. However, providing several representations does not ensure that learners can use them together. Research in chemistry education has demonstrated differences between experts and novices in interpreting and coordinating representations of scientific phenomena~\citep{kozma1997multimedia}. Rau~\citep{rau2017representations} further identifies conditions under which multiple visual representations can support STEM learning.

A central challenge is recognizing correspondences between representations that communicate different aspects of the same system. Gentner's structure-mapping theory provides a foundation for understanding relational correspondence, emphasizing relationships among elements rather than similarities in their appearance~\citep{gentner1983mapping}. For example, a three-dimensional structural model and a force diagram may represent the same assembly while emphasizing different properties. Understanding how they correspond requires learners to connect spatial arrangements with the relationships expressed through the diagram.
Chemistry education provides a particularly relevant example of this challenge. The distinction among macroscopic, submicroscopic, and symbolic representations has highlighted the importance of connecting observable phenomena, underlying mechanisms, and formal descriptions~\citep{taber2013triplet}. These representations may describe different aspects of a phenomenon without necessarily corresponding to distinct physical scales.

For cross-scale learning, the instructional value of multiple representations lies in helping learners establish relationships that cannot be understood through a single view. The challenge is not the number of representations available, but whether learners can recognize their correspondences, combine the information they provide, and use it to construct meaningful explanations.

\subsection{Systems Thinking and Causal Reasoning}
\label{sec:systems-thinking}

Systems thinking concerns how components, interactions, and patterns of organization contribute to the behavior of a larger system. It requires learners to examine relationships rather than treat individual elements as independent. Jacobson and Wilensky~\citep{jacobson2006complex} describe the importance of complex-systems perspectives in science education, particularly for understanding interactions across organizational levels and the emergence of system-level behavior.
Understanding these interactions also requires learners to distinguish the structures that constitute a system from the mechanisms through which it operates. The Structure--Behavior--Function framework provides an account of these relationships by connecting system components, their behaviors, and their functions. Hmelo-Silver and Pfeffer~\citep{hmelosilver2004sbf} found that experts and novices differ in how they explain complex systems, particularly in their attention to mechanisms and functional relationships. Related research in biology further demonstrates the importance of understanding how system structures and processes interact rather than identifying visible components alone~\citep{hmelosilver2007fish}.

Causal reasoning becomes particularly important when an observed outcome must be explained through processes operating at another level. For example, local material interactions during manufacturing can influence the properties of a finished component, while changes in molecular interactions can affect cellular behavior. In each case, a meaningful explanation requires learners to connect the relevant structures and mechanisms to the resulting outcome.
These relationships also support reasoning beyond an observed example. Goldstone and Wilensky~\citep{goldstone2008transfer} investigated how grounded representations and computational models can support the learning and transfer of complex-systems principles. Their work highlights the importance of understanding relationships that can be applied to new situations, rather than relying exclusively on the features of a familiar example.

Together, systems thinking and causal reasoning provide a foundation for explaining how interactions at one level contribute to behavior at another. For CSSR, these explanations are particularly relevant when the relationships involve spatial structures, physical arrangements, or representations distributed across levels.

\subsection{Embodied and Immersive Learning}
\label{sec:immersive-learning}

Immersive technologies provide opportunities to explore scientific and engineering systems through interactive spatial representations. Virtual reality (VR) can make structures and environments accessible at scales that are difficult to experience directly, while augmented reality (AR) can connect digital representations with physical objects and activities. These capabilities allow learners to inspect spatial relationships, manipulate representations, and investigate processes that may otherwise be difficult to observe.
Research on embodied learning further suggests that physical interaction can support conceptual understanding when actions are meaningfully connected to the concepts being learned. Lindgren and colleagues~\citep{lindgren2016embodied}, for example, investigated embodied interaction in a mixed-reality physics simulation and demonstrated its potential to support learning and engagement.

The Cognitive Affective Model of Immersive Learning (CAMIL) explains how immersive characteristics may influence learning through cognitive and affective processes, including presence, agency, and cognitive load~\citep{makransky2021camil}. Petersen, Petkakis, and Makransky~\citep{petersen2022interactivity} further investigated the contributions of immersion and interactivity to learning in VR. These perspectives emphasize that the educational value of immersive environments depends on how their capabilities engage learners in meaningful activities.
Existing immersive learning systems illustrate these opportunities across scientific and engineering domains. VR applications have supported the exploration of cellular structures~\citep{laubscher2024cell}, while immersive manufacturing environments have provided opportunities to practice machining and additive manufacturing operations~\citep{conesa2024lathe,ahmed2026additive}. Other systems have supported microfabrication training and the visualization of electrical circuit principles~\citep{wang2024cleanroom,wolbach2024circuits}. In structural engineering, AR and VR applications have enabled learners to investigate structural relationships and system behavior~\citep{turkan2017structural,rodgers2026building}.

These applications demonstrate how immersive technologies can make complex information accessible through spatial exploration and interaction. However, the learning benefits of immersion are not consistent across instructional settings. Makransky, Terkildsen, and Mayer~\citep{makransky2019presence} found that an immersive laboratory simulation increased presence but resulted in lower learning outcomes than a desktop version under the conditions examined. Parong and Mayer~\citep{parong2018science,parong2021cognitive} investigated cognitive and affective processes involved in immersive science learning, while reviews and meta-analyses have identified variation in learning outcomes across technologies, instructional designs, and educational contexts~\citep{hamilton2021review,coban2022meta}.

These findings reinforce the importance of instructional design. Cognitive load theory explains how unnecessary processing demands may interfere with learning~\citep{sweller1988cognitive}, while multimedia learning research provides guidance on coordinating information and directing learners' attention toward relevant content~\citep{mayer2003cognitiveload}. In cross-scale learning activities, these concerns become particularly important when learners must navigate between levels, coordinate representations, and interpret system behavior while managing the interactions required by an immersive environment.

Immersive technologies therefore provide promising mechanisms for supporting cross-scale learning, but their capabilities must be connected to explicit reasoning objectives. The instructional question is not simply how learners can explore a complex system, but how that exploration can help them understand the relationships that explain its behavior.

\subsection{Synthesis: The Challenge of Cross-Scale Integration}
\label{sec:theoretical-synthesis}

The research reviewed in this section provides complementary foundations for understanding complex scientific and engineering systems. Spatial cognition addresses the interpretation of spatial structures; size-and-scale cognition concerns magnitude and proportional relationships; multiple-representation research examines how learners coordinate different forms of information; and systems thinking explains how interactions among components contribute to larger-scale behavior. Research on immersive learning identifies technological and instructional opportunities for engaging with these forms of information.

A common challenge emerges when a learning task requires these forms of reasoning to operate together. Learners may understand relevant structures, representations, or scientific principles independently, yet struggle to establish the relationships needed to explain how a system behaves across levels. The difficulty may involve identifying the relevant levels, connecting corresponding structures, combining information from different representations, or explaining how local interactions contribute to broader outcomes.
We refer to the coordination of these demands as \textit{cross-scale integration}. Existing research provides important explanations of its constituent processes, but there remains a need for an instructional account that identifies how these processes work together within cross-scale learning tasks and connects them to the design and assessment of learning environments.

CSSR addresses this need by organizing the reasoning demands involved in establishing and using relationships across physical scales or organizational levels. The following section defines the framework, characterizes its constituent processes, and establishes its conceptual boundaries.

%% file: sections/03-Concept.tex
\section{Conceptualizing Cross-Scale Spatial Reasoning}
\label{sec:conceptualizing-cssr}

The theoretical foundations reviewed in Section~\ref{sec:theoretical-foundations} highlight a central challenge in learning about complex systems: understanding individual structures, processes, or representations does not necessarily enable learners to explain the relationships connecting them across levels. This section defines Cross-Scale Spatial Reasoning (CSSR), identifies five interrelated reasoning processes, and examines how they contribute to understanding system behavior.

\subsection{Defining CSSR}
\label{sec:cssr-definition}

We define \textit{Cross-Scale Spatial Reasoning} (CSSR) as the coordinated reasoning involved in establishing and using spatial and structural relationships across physical scales or organizational levels to explain system behavior and predict the consequences of changes.
This definition emphasizes three related demands. Learners must identify the levels relevant to a problem and establish the relationships connecting them. They may also need to interpret these relationships across different representations, each revealing particular aspects of the system. Finally, they must use the relationships to construct explanations, evaluate possible changes, or predict outcomes. Neither recognizing structures at multiple levels nor moving between representations is sufficient without meaningful integration.
CSSR can involve changes in physical scale, organizational level, or both. Molecular interactions and cellular processes, for example, involve relationships across physical scales and levels of organization. The relationship between a structural connection and the behavior of an entire assembly may primarily concern organizational levels. Changes in representation, such as moving between a three-dimensional model and a functional diagram, may also be involved without necessarily changing the underlying scale or organizational level.
The defining characteristic of CSSR is therefore \textit{cross-scale integration}: using relationships between levels to construct a coherent explanation of how a system works. A learner may understand individual components and processes yet struggle to explain how their interactions contribute to larger-scale behavior. CSSR makes this distinction an explicit focus of learning and assessment.

\subsection{Illustrative Example: Diagnosing Cross-Scale Reasoning Difficulties}
\label{sec:cssr-worked-example}

Consider a learner investigating why a digital circuit exhibits an unexpected signal delay. The learner can identify its components, interpret a circuit schematic, and inspect the physical layout. Nevertheless, these abilities alone may not explain why the observed delay occurs.
A complete explanation requires the learner to identify the relevant signal path, connect the schematic to its physical implementation, and consider how device and interconnect characteristics contribute to timing. The learner must combine information distributed across representations and organizational levels to explain the observed behavior.
Different learners may encounter difficulties at different points. One may fail to identify which levels are relevant to the timing problem. Another may understand the schematic and layout independently but struggle to establish correspondences between them. A third may identify the relevant structures yet be unable to explain how their properties contribute to signal delay.
These difficulties may produce similar incorrect answers, but they reflect different reasoning demands. Distinguishing them matters because the instructional support required to identify relevant levels may differ from the support needed to establish relationships or construct causal explanations.
A further test of understanding would involve presenting an unfamiliar layout modification. Rather than simply predicting whether the delay increases or decreases, the learner would need to justify the prediction using the physical relationships and mechanisms involved, while accounting for relevant constraints.
This example illustrates the purpose of CSSR: to characterize the reasoning involved in connecting information across levels and distinguish difficulties that might otherwise be treated as a general lack of understanding. Although illustrated here through circuit design, the same distinctions can inform the analysis of tasks involving biological processes, manufacturing operations, and structural systems.

\subsection{Five Constituent Processes}
\label{sec:cssr-processes}

We propose five interrelated processes that characterize cross-scale integration: \textit{scale navigation}, \textit{cross-scale relational mapping}, \textit{representational coordination}, \textit{dynamic and causal reasoning}, and \textit{consequence-oriented reasoning}. These processes draw on established research in spatial cognition, relational mapping, multiple representations, and systems understanding~\citep{gentner1983mapping,ainsworth2006deft,wilensky1999levels,hmelosilver2004sbf}.
The processes describe distinguishable reasoning demands rather than independent cognitive abilities or stages in a fixed sequence. Their relative importance depends on the task, the information available, and the learner's prior knowledge.

\subsubsection{Scale Navigation}

Scale navigation involves identifying which physical scales or organizational levels are relevant to a problem, determining what information each provides, and deciding when another level must be examined. Its central challenge is cognitive rather than navigational. A learner may move between magnifications or views without recognizing which level contains the information needed for an explanation. Effective scale navigation requires understanding why a particular level is relevant, what it reveals, and what remains unexplained.

\subsubsection{Cross-Scale Relational Mapping}

Cross-scale relational mapping involves establishing meaningful connections among structures and processes at different levels. These connections may involve part--whole relationships, spatial arrangements, functional dependencies, or interactions between components and larger systems. The emphasis is on relationships within the system rather than the appearance of individual structures. Identifying corresponding components across levels is not sufficient unless the learner can explain how they are connected and why those connections matter. This process draws on structure-mapping research, which emphasizes relational correspondences rather than superficial similarities~\citep{gentner1983mapping}.

\subsubsection{Representational Coordination}

Representational coordination concerns interpreting and combining information from different representations of a system, including spatial models, functional diagrams, quantitative descriptions, and simulations. Because each representation emphasizes particular properties, learners must identify correspondences and determine how the information contributes to a shared explanation. This may involve connecting geometry with functional relationships or interpreting a physical configuration alongside its modeled behavior.
Representational coordination differs from relational mapping in its emphasis. Relational mapping concerns connections among structures and processes across levels, whereas representational coordination concerns how those connections are communicated and interpreted through different forms of information. A learner may coordinate two representations at the same organizational level without engaging in cross-scale reasoning. Conversely, cross-scale relationships may be understood through a single integrated representation. The two processes often support one another, but they are not equivalent~\citep{ainsworth2006deft,rau2017representations}.

\subsubsection{Dynamic and Causal Reasoning}

Dynamic and causal reasoning involves explaining how interactions, changes, and underlying mechanisms produce system behavior. Learners must move beyond recognizing that an outcome occurs to explaining why it occurs and how processes at different levels contribute to it. Such explanations may involve tracing the effects of local interactions, identifying dependencies, or examining how behavior develops over time. The objective is to connect relevant mechanisms to observed outcomes rather than infer causation from association alone. This process builds on research examining how learners understand structures, behaviors, and functions within complex systems~\citep{hmelosilver2004sbf,hmelosilver2007fish}.

\subsubsection{Consequence-Oriented Reasoning}

Consequence-oriented reasoning involves using an understanding of cross-scale relationships to predict and evaluate possible outcomes. Learners consider how changes in structures, conditions, or interactions at one level may influence other levels of the system. This process requires more than predicting an outcome: learners must justify their predictions using relevant mechanisms and consider how competing effects, constraints, or uncertainty may influence the result. Consequence-oriented reasoning is particularly important in inquiry and design tasks, where learners evaluate alternative configurations or explain the effects of unfamiliar changes. It also applies to scientific systems that are investigated rather than deliberately modified.

\subsection{How the Processes Work Together}
\label{sec:cssr-process-integration}

\begin{figure*}[t]
\centering

\resizebox{0.96\textwidth}{!}{%
\begin{tikzpicture}[
    font=\footnotesize,
    process/.style={
        draw=black!35,
        line width=0.5pt,
        rounded corners=2pt,
        fill=black!2,
        text width=4.25cm,
        minimum height=2.05cm,
        inner sep=7pt,
        align=left
    },
    centerbox/.style={
        draw=blue!45!black,
        line width=0.7pt,
        rounded corners=3pt,
        fill=blue!5,
        text width=4.9cm,
        minimum height=1.35cm,
        inner sep=8pt,
        align=center
    },
    link/.style={
        draw=black!40,
        line width=0.7pt,
        <->,
        >=Latex
    }
]

% Top row
\node[process] (navigation) at (-4.8,2.9) {
    {\bfseries Scale navigation}\\[5pt]
    Which scales or levels are relevant?\\[2pt]
    What does each level reveal or omit?\\[2pt]
    When is another level needed?
};

\node[process] (mapping) at (0,2.9) {
    {\bfseries Cross-scale relational mapping}\\[5pt]
    How are structures connected across levels?\\[2pt]
    Which relationships remain meaningful?\\[2pt]
    How do local interactions relate to the whole?
};

\node[process] (representation) at (4.8,2.9) {
    {\bfseries Representational coordination}\\[5pt]
    What does each representation communicate?\\[2pt]
    How do corresponding elements align?\\[2pt]
    What information must be combined?
};

% Center
\node[centerbox] (integration) at (0,0) {
    {\bfseries\small Cross-scale integration}\\[6pt]
    Connect spatial and structural relationships
    across levels to explain system behavior
    and predict consequences.
};

% Bottom row
\node[process] (causal) at (-2.45,-2.9) {
    {\bfseries Dynamic and causal reasoning}\\[5pt]
    How do interactions produce behavior?\\[2pt]
    Why does a particular outcome occur?\\[2pt]
    How do effects develop across levels?
};

\node[process] (consequence) at (2.45,-2.9) {
    {\bfseries Consequence-oriented reasoning}\\[5pt]
    What changes when conditions are modified?\\[2pt]
    How might effects propagate across levels?\\[2pt]
    What outcomes or trade-offs are expected?
};

% Bidirectional connections
\draw[link]
    (navigation.south east) --
    (integration.north west);

\draw[link]
    (mapping.south) --
    (integration.north);

\draw[link]
    (representation.south west) --
    (integration.north east);

\draw[link]
    (integration.south west) --
    (causal.north);

\draw[link]
    (integration.south east) --
    (consequence.north);

\draw[link]
    (causal.east) --
    (consequence.west);

\end{tikzpicture}%
}

\caption{
Five interrelated processes of Cross-Scale Spatial
Reasoning (CSSR) and the guiding questions associated
with each. Together, these processes support the
integration of spatial and structural relationships
across levels to explain system behavior and predict
consequences. The bidirectional connections emphasize
their interdependence rather than a fixed reasoning
sequence.
}
\label{fig:cssr-process-model}

\end{figure*}
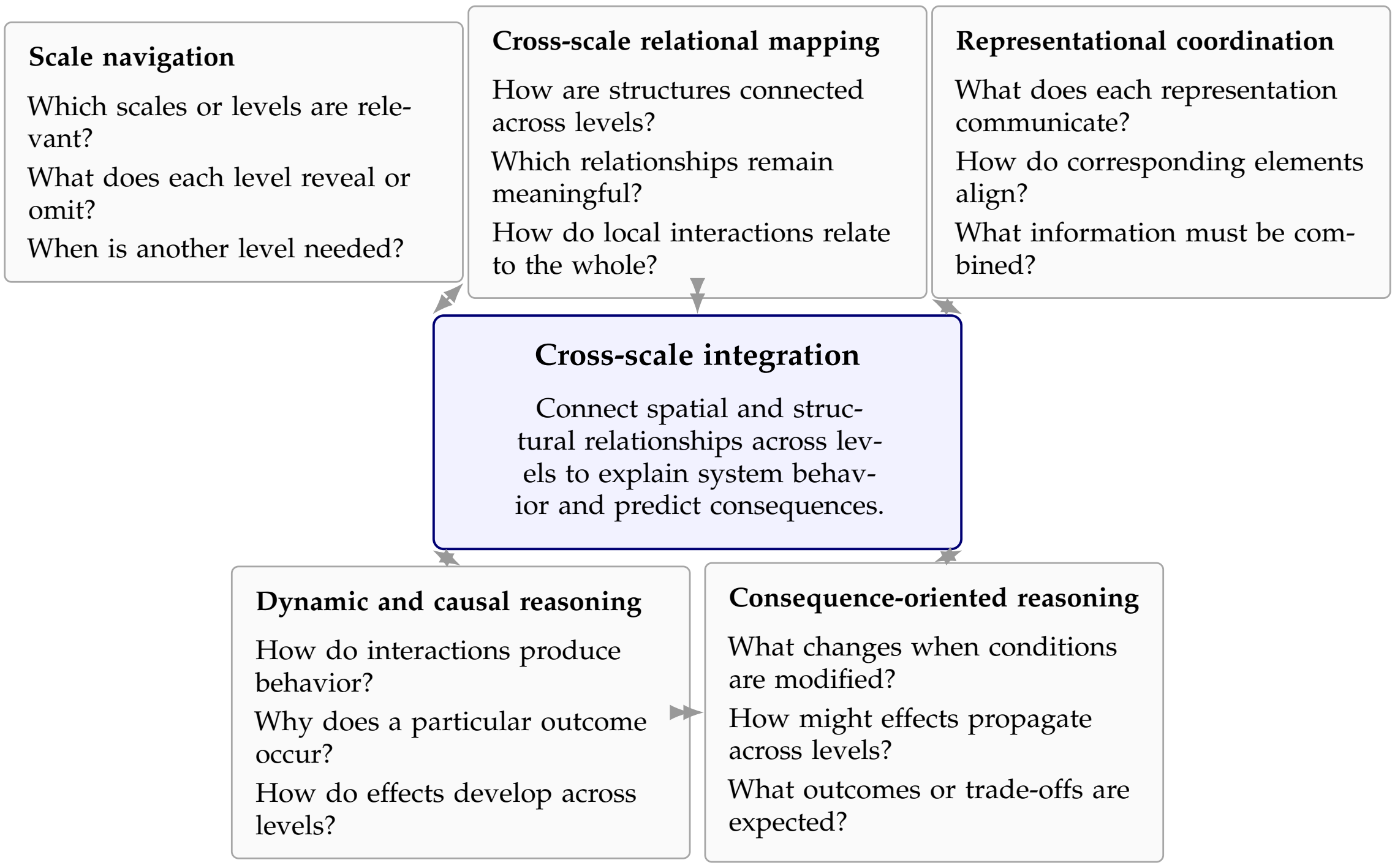

The five CSSR processes describe complementary aspects of cross-scale understanding. Scale navigation helps learners identify relevant levels, relational mapping establishes connections among structures, and representational coordination supports the interpretation of those connections across different forms of information. Dynamic and causal reasoning draws on these relationships to explain system behavior, while consequence-oriented reasoning uses them to evaluate possible outcomes.
These processes are interdependent and need not occur sequentially. A learner may begin with an unexpected outcome, identify a possible mechanism, examine another organizational level, and revise an earlier explanation. Cross-scale reasoning may therefore involve repeated movement between levels, representations, explanations, and predictions as new questions arise.
Figure~\ref{fig:cssr-process-model} summarizes the five processes and the guiding questions associated with each. The diagram emphasizes their complementary roles in cross-scale integration without prescribing a fixed sequence or limiting the framework to a particular disciplinary context.

% INSERT YOUR EXISTING FIGURE 2 TIKZ CODE HERE.
% Retain \label{fig:cssr-process-model} in the figure environment.

The framework also helps distinguish different learning difficulties. A learner may identify relevant levels but fail to establish the relationships between them. Another may coordinate representations accurately yet construct an incomplete causal explanation. These differences can inform the selection of instructional support and provide a basis for evaluating specific aspects of cross-scale understanding.

\subsection{Conceptual Boundaries}
\label{sec:cssr-boundaries}

CSSR draws on established forms of reasoning but is not synonymous with any one of them. Spatial visualization may involve interpreting the geometry of a structure without examining its relationships to another level. Size-and-scale reasoning may involve comparing magnitudes or proportions without explaining system behavior~\citep{magana2012fs2c,cheek2017scale}. Representational competence may involve translating between different forms of information at the same organizational level. These capabilities can contribute to CSSR, but none independently establishes cross-scale integration.
Systems thinking is particularly closely related because it concerns relationships among components and larger-scale behavior~\citep{wilensky1999levels,jacobson2006complex}. CSSR focuses more specifically on tasks in which spatial or structural relationships across physical scales or organizational levels are central to the explanation. Not every systems-thinking task is spatial, and not every cross-scale relationship involves emergent behavior.
The proposed contribution of CSSR is therefore not the identification of entirely new cognitive abilities. It is an integrated account of how established forms of reasoning contribute to a particular class of learning tasks. Whether this account provides additional explanatory value or constitutes a distinguishable cognitive construct remains an empirical question, addressed in Section~\ref{sec:cssr-construct-validity}.

\subsection{CSSR as a Learning Objective}
\label{sec:cssr-operationalization}

For CSSR to guide instructional design, the proposed processes must be connected to observable learning outcomes. Successful navigation or manipulation of representations may demonstrate interaction proficiency, but does not by itself establish cross-scale understanding. More direct evidence comes from learners' ability to identify relevant levels, establish relationships across them, combine information from different representations when necessary, explain system behavior, and justify predictions about unfamiliar changes.
These outcomes can be examined through written explanations, annotated diagrams, structured comparisons, and interactive problem-solving tasks. Their interpretation must account for learners' prior domain knowledge and understanding of the individual representations involved.

The next section translates these reasoning objectives into a learning-centered framework for designing immersive environments, while Section~\ref{sec:cssr-assessment-validation} examines how cross-scale understanding can be assessed and validated.

%% file: sections/04-Framework.tex
\section{A Learning-Centered Framework for Immersive Learning Design}
\label{sec:cssr-design-framework}

Designing immersive learning environments to support CSSR requires more than making structures visible at different scales. Learners need opportunities to establish relationships across levels, explain the mechanisms connecting them, and apply those explanations to new problems. The central design challenge is therefore to determine which forms of cross-scale reasoning a learning task requires and how immersive interactions and instructional support can help learners develop them.
We propose a learning-centered framework that connects the five CSSR processes to immersive design mechanisms, instructional scaffolding, and assessment. Rather than beginning with a technology or interaction technique, the framework begins with a cross-scale learning objective. It then identifies the relationships learners must understand, the information and representations needed to investigate them, the difficulties learners may encounter, and the evidence required to demonstrate understanding.

\subsection{From Learning Objectives to Design Decisions}
\label{sec:cssr-design-approach}

A CSSR-centered learning activity begins with a problem that requires learners to explain relationships across physical scales or organizational levels. Designers must identify the structures, processes, and relationships relevant to the problem, establish what learners already understand, and determine which aspects of cross-scale reasoning the task demands. These demands may involve selecting relevant levels, mapping relationships among structures, coordinating representations, explaining mechanisms, or predicting the consequences of changes.
The circuit-delay example from Section~\ref{sec:cssr-worked-example} illustrates this approach. The learning objective is not simply to inspect a circuit schematic or explore its physical layout, but to explain how properties of the physical implementation contribute to circuit timing. Achieving this objective requires learners to identify the relevant levels, connect signal paths across representations, examine device and interconnect characteristics, and explain their contribution to the observed delay. Learners should also be able to use these relationships to justify predictions about unfamiliar configurations.
The framework translates such objectives into four connected design decisions. First, designers specify the cross-scale relationships learners need to understand and the reasoning processes involved. Second, they select representations and interaction mechanisms that make those relationships available for investigation. Third, they provide instructional support aligned with the difficulties learners may encounter. Finally, they assess whether learners can explain and apply the relationships independently, rather than merely complete the interactions provided by the environment. These decisions are interdependent and may be revisited as the learning activity is developed and evaluated.

\subsection{Design Mechanisms}
\label{sec:cssr-design-mechanisms}

The framework identifies four broad mechanisms that may support cross-scale learning: navigation with relational context, coordinated representations, interactive manipulation and simulation, and contextual feedback. These mechanisms are not unique to CSSR and have been used in different forms within existing immersive learning systems. Their role in the proposed framework is to support specific reasoning demands rather than serve as educational objectives in themselves. Table~\ref{tbl:cssr-process-design-mapping} illustrates how the five CSSR processes can inform the selection of instructional mechanisms and corresponding evidence of learning.

\subsubsection{Navigation with Relational Context}

Immersive environments allow learners to examine structures at different magnifications, spatial scales, or organizational levels. However, changing the level of examination can obscure how a structure relates to the larger system. Navigation should therefore preserve relevant context, helping learners recognize where a structure belongs, what information the new level provides, and why examining it is necessary for the learning task.
For example, moving from a cellular view to a molecular model may help learners examine the structure of a protein. This transition becomes more useful for cross-scale reasoning when the protein remains connected to its role in the cellular pathway. Similarly, examining a structural connection in architecture should preserve its relationship to the larger assembly. Linked highlighting, contextual landmarks, and persistent relational cues can help learners maintain these connections during exploration. The objective is not simply to enable movement between levels, but to support purposeful scale navigation and relational understanding.

\subsubsection{Coordinated Representations}

Many cross-scale explanations require learners to combine information from representations that emphasize different properties of a system. A three-dimensional model may communicate spatial geometry, a diagram may describe functional relationships, and a simulation may show behavior over time. Coordinating these representations can make relevant correspondences accessible and reduce the need for learners to reconstruct them entirely through mental comparison.
In semiconductor design, synchronized schematic and layout views could help learners trace a signal path and relate its logical connectivity to its physical implementation. In structural engineering, a three-dimensional model could be coordinated with force diagrams and representations of load distribution. Such mechanisms should help learners recognize what corresponds across views, what information each representation contributes, and how that information supports an explanation. Their design must also account for learners' familiarity with the representations and avoid introducing unnecessary visual or interaction complexity~\citep{ainsworth2006deft,rau2017representations}.

\subsubsection{Interactive Manipulation and Simulation}

Interactive manipulation and simulation provide opportunities to investigate how changes in structures, conditions, or interactions influence system behavior. Learners may modify a component, process parameter, or configuration and observe the resulting response. However, observing a change is not equivalent to explaining it. CSSR-centered activities should require learners to connect the simulated outcome to the relationships and mechanisms responsible for it.
In manufacturing, learners might modify cutting conditions and inspect changes in modeled tool--material interactions before relating them to the characteristics of the finished component. In biology, a modeled modification to a protein could be used to investigate possible changes in molecular interactions and their consequences within a cellular pathway. Prediction prompts before manipulation and explanation tasks afterward can encourage learners to use the simulation as evidence for causal reasoning rather than rely on trial and error. Because simulations necessarily simplify real systems, designers should also communicate their assumptions, constraints, and limitations.

\subsubsection{Contextual Feedback}

Contextual feedback can help learners recognize relationships they have overlooked, reconsider incomplete explanations, or identify information needed to resolve an uncertainty. Rather than providing a complete solution, feedback can direct attention toward relevant structures, suggest another representation, or prompt learners to explain a causal connection.
For example, a learner investigating circuit delay may concentrate on the schematic without considering the physical routing of interconnections. Feedback could highlight the corresponding path in the layout and ask the learner to explain its possible contribution to timing. A different learner who has correctly identified the path but constructed an incomplete explanation may instead benefit from a prompt concerning device or interconnect properties. The appropriate feedback therefore depends on the underlying reasoning difficulty, not simply on whether the learner has performed a particular interaction.

\subsection{Scaffolding Cross-Scale Reasoning}
\label{sec:cssr-scaffolding}

Learners bring different levels of prior knowledge and experience to complex-system tasks. Some may struggle to identify which levels are relevant, while others may understand individual representations but fail to establish relationships across them. Still others may recognize the relevant connections without being able to explain how they contribute to system behavior. Instructional scaffolding should therefore respond to the specific reasoning demands that learners have difficulty addressing.
One approach is to introduce cross-scale relationships progressively. Learners may first examine components at one level and then investigate their relationships to a larger system. Alternatively, an activity may begin with an unexplained system-level outcome and guide learners toward the smaller-scale interactions responsible. Neither direction is inherently preferable; the choice depends on the learning objective, the characteristics of the system, and learners' prior knowledge.
Prediction and explanation prompts offer another form of support. Before observing a simulation, learners may be asked to anticipate an outcome and justify their prediction through relevant relationships. After investigating the result, they can compare the outcome with their expectations and revise their explanation. Such activities draw on established research on self-explanation and generative learning~\citep{chi1989selfexplanations,fiorella2015generative}.
Scaffolding should become less explicit as learners develop understanding. Early activities may provide correspondence cues, linked highlighting, or guided explanations, while later activities require learners to identify relevant relationships and construct explanations more independently. This approach is consistent with research on scaffolded inquiry and fading guidance~\citep{quintana2004scaffolding,renkl2003fading}. The aim is to help learners develop the reasoning needed to understand a system, rather than have the environment perform that reasoning for them.
Adaptive scaffolding may provide further opportunities to personalize support, but it requires reliable evidence of learners' difficulties. Repeated navigation, extended task time, or frequent interaction with a representation does not necessarily indicate confusion. These behaviors may also reflect deliberate exploration or careful comparison. Where possible, adaptive decisions should therefore draw on learners' explanations, predictions, and demonstrated understanding, alongside interaction data, rather than infer cognitive difficulties from interaction traces alone.

\subsection{An Actionable Design Specification}
\label{sec:cssr-design-specification}

The proposed framework can inform both general instructional decisions and the design of activities within particular disciplines. Table~\ref{tbl:cssr-process-design-mapping} maps the five CSSR reasoning processes to possible immersive design mechanisms and observable evidence of learning. The mapping is illustrative rather than prescriptive: individual mechanisms may support several processes, and the appropriate combination depends on the learning task.

\begin{table*}[t]
\centering
\caption{Illustrative relationships between CSSR reasoning processes, immersive design mechanisms, and evidence of learning.}
\label{tbl:cssr-process-design-mapping}
\footnotesize
\setlength{\tabcolsep}{6pt}
\renewcommand{\arraystretch}{1.35}
\begin{tabularx}{\textwidth}{@{} >{\raggedright\arraybackslash}p{0.24\textwidth} >{\raggedright\arraybackslash}X >{\raggedright\arraybackslash}X @{}}
\toprule
\textbf{CSSR process} & \textbf{Possible design mechanisms} & \textbf{Evidence of learning} \\
\midrule
\textbf{Scale navigation} & Context-preserving transitions, scale indicators, and structural landmarks. & Selecting relevant levels and explaining why each is needed. \\
\midrule[0.25pt]
\textbf{Cross-scale relational mapping} & Linked structural cues, part--whole highlighting, and relational annotations. & Identifying and explaining relationships connecting structures across levels. \\
\midrule[0.25pt]
\textbf{Representational coordination} & Synchronized views, correspondence cues, and complementary spatial and symbolic representations. & Integrating information from different representations into a coherent explanation. \\
\midrule[0.25pt]
\textbf{Dynamic and causal reasoning} & Interactive simulations, temporal visualizations, and mechanism-focused explanation prompts. & Explaining how interactions and changes produce system behavior. \\
\midrule[0.25pt]
\textbf{Consequence-oriented reasoning} & What-if exploration, parameter manipulation, and prediction--observation comparisons. & Justifying predicted outcomes of unfamiliar changes using cross-scale relationships. \\
\bottomrule
\end{tabularx}
\end{table*}

The circuit-delay example provides a more specific application. Table~\ref{tbl:cssr-design-specification} translates a cross-scale learning objective into representations, immersive interactions, instructional scaffolding, and assessment evidence. It illustrates how the general reasoning processes can guide decisions about the content and structure of a particular learning activity.

% KEEP YOUR EXISTING CIRCUIT-DELAY DESIGN SPECIFICATION TABLE HERE.

\begin{table*}[t]
\centering
\caption{Illustrative design specification for a CSSR-centered learning activity investigating circuit delay.}
\label{tbl:cssr-design-specification}
\footnotesize
\setlength{\tabcolsep}{9pt}
\renewcommand{\arraystretch}{1.45}
\begin{tabularx}{\textwidth}{@{}>{\columncolor{blue!4}\raggedright\arraybackslash}p{0.24\textwidth}>{\raggedright\arraybackslash}X@{}}
\toprule
\rowcolor{blue!10}
\textbf{Design element} & \textbf{Learning activity specification} \\
\midrule
\textbf{Learning objective} & Explain how the physical implementation of a circuit influences signal delay. \\
\midrule[0.25pt]
\textbf{Relevant levels} & Device properties, physical interconnections, and circuit-level behavior. \\
\midrule[0.25pt]
\textbf{Reasoning challenge} & Connect signal paths in a schematic to their physical implementation and resulting timing behavior. \\
\midrule[0.25pt]
\textbf{Representations} & Circuit schematic, physical layout, device and interconnect information, and timing results. \\
\midrule[0.25pt]
\textbf{Immersive interactions} & Link corresponding elements across representations, inspect interconnections, and compare alternative layouts. \\
\midrule[0.25pt]
\textbf{Instructional scaffolding} & Provide correspondence cues, prediction prompts, and feedback on incomplete causal explanations. \\
\midrule[0.25pt]
\textbf{Learning assessment} & Explain observed delay and justify predictions for an unfamiliar layout modification. \\
\bottomrule
\end{tabularx}
\end{table*}

The specification illustrates a distinction between technology-centered and learning-centered approaches. A technology-centered activity might emphasize the ability to navigate a three-dimensional circuit or manipulate its components. A CSSR-centered activity instead asks whether those interactions help learners connect relevant information, construct an accurate explanation of circuit behavior, and justify predictions about changes in its physical implementation.
The same approach can guide other disciplines without prescribing identical learning activities. Designers can replace the representations, scientific mechanisms, and instructional interactions while retaining the central questions: which cross-scale relationships must learners understand, what reasoning difficulties might prevent them from establishing those relationships, which forms of support could help, and what evidence would demonstrate understanding?

\subsection{Design Principles and Research Propositions}
\label{sec:cssr-design-principles}

The framework suggests six principles for designing learning activities that support CSSR:

\begin{enumerate}
\item \textbf{Begin with a cross-scale learning objective.} Specify the relationships learners must explain and the reasoning processes involved before selecting immersive features.
\item \textbf{Preserve context across levels.} Maintain meaningful structural and functional connections as learners move between scales or organizational levels.
\item \textbf{Coordinate complementary representations.} Make relevant correspondences accessible when different representations contribute to the same explanation.
\item \textbf{Require prediction and explanation.} Encourage learners to investigate mechanisms, anticipate outcomes, and justify results using cross-scale relationships.
\item \textbf{Provide targeted and fading support.} Address identified reasoning difficulties while progressively increasing learners' independence.
\item \textbf{Assess cross-scale understanding directly.} Evaluate explanations and justified predictions rather than infer learning from navigation, engagement, or task completion alone.
\end{enumerate}

These principles build on established research on multiple representations, scaffolded inquiry, and generative learning. Their proposed contribution lies in organizing these instructional approaches around the reasoning demands of cross-scale tasks and connecting them to observable learning outcomes. The following research propositions specify conditions under which particular design mechanisms may support this understanding.

\subsubsection{Research Propositions}
\label{sec:cssr-propositions}

\textbf{P1: Relational continuity.} For tasks requiring transitions between physical scales or organizational levels, representations that preserve relevant structural relationships will improve learners' ability to reconstruct cross-scale connections compared with equivalent views presented without explicit relational continuity. The benefit is expected to be greatest when learners understand the individual levels but struggle to establish relationships between them.
\textbf{P2: Coordinated representations.} When a cross-scale explanation requires information distributed across multiple representations, linked views with meaningful correspondence cues will support more accurate relational mapping and integration than independently presented views containing equivalent information. The benefit is expected to depend on learners having sufficient knowledge to interpret the individual representations.
\textbf{P3: Prediction and explanation.} When a learning task requires learners to explain how changes at one level influence behavior at another, activities combining prediction, interactive investigation, and causal explanation will support stronger cross-scale understanding than observation-focused activities containing the same scientific information. These benefits are expected to be particularly relevant when outcomes cannot be adequately explained through inspection of individual structures alone.
\textbf{P4: Targeted scaffolding.} Instructional guidance matched to an identified cross-scale reasoning difficulty will improve performance on the corresponding task relative to general guidance of comparable detail. As learners develop understanding, progressively reducing support is expected to encourage more independent identification, explanation, and application of cross-scale relationships.
\textbf{P5: Transfer.} Instruction that explicitly develops relationships connecting physical scales or organizational levels will support stronger performance on unfamiliar but structurally related problems than instruction emphasizing individual representations alone. This benefit is expected when the new problem preserves relevant relational principles and learners possess sufficient domain knowledge to interpret its structures and mechanisms.

These propositions concern the effects of particular instructional mechanisms rather than the benefits of immersive presentation alone. Their evaluation requires comparisons that account for scientific content, prior knowledge, task difficulty, and the information available across experimental conditions. Studies comparing immersive and non-immersive implementations may further clarify when the affordances of AR and VR provide additional value for cross-scale learning.
Taken together, the framework provides a systematic basis for connecting CSSR reasoning objectives to instructional decisions. Designers identify the relationships learners need to understand, select interactions that support their investigation, provide guidance responsive to learners' difficulties, and assess whether learners can explain and apply those relationships independently.

%% file: sections/05-Applications.tex
\section{Applying CSSR to Immersive Learning Across Domains}
\label{sec:cssr-applications}

The CSSR framework can inform the design of immersive learning activities across disciplines in which understanding requires learners to establish relationships between physical scales or organizational levels. Although the underlying structures, mechanisms, and representations differ across domains, the instructional challenge is similar: learners must identify relevant levels, establish meaningful connections, coordinate information, and use these relationships to explain system behavior or predict consequences.
This section illustrates how the framework can be translated into learning objectives, interactions, and assessment tasks in biology, manufacturing, semiconductor design, and architecture. These examples are proposed instructional scenarios rather than implemented or empirically evaluated CSSR interventions. Existing immersive learning systems provide relevant technological and educational precedents, but their reported outcomes do not establish the effectiveness of the proposed framework.

\begin{figure}[t]
    \centering
    \includegraphics[width=0.95\linewidth]{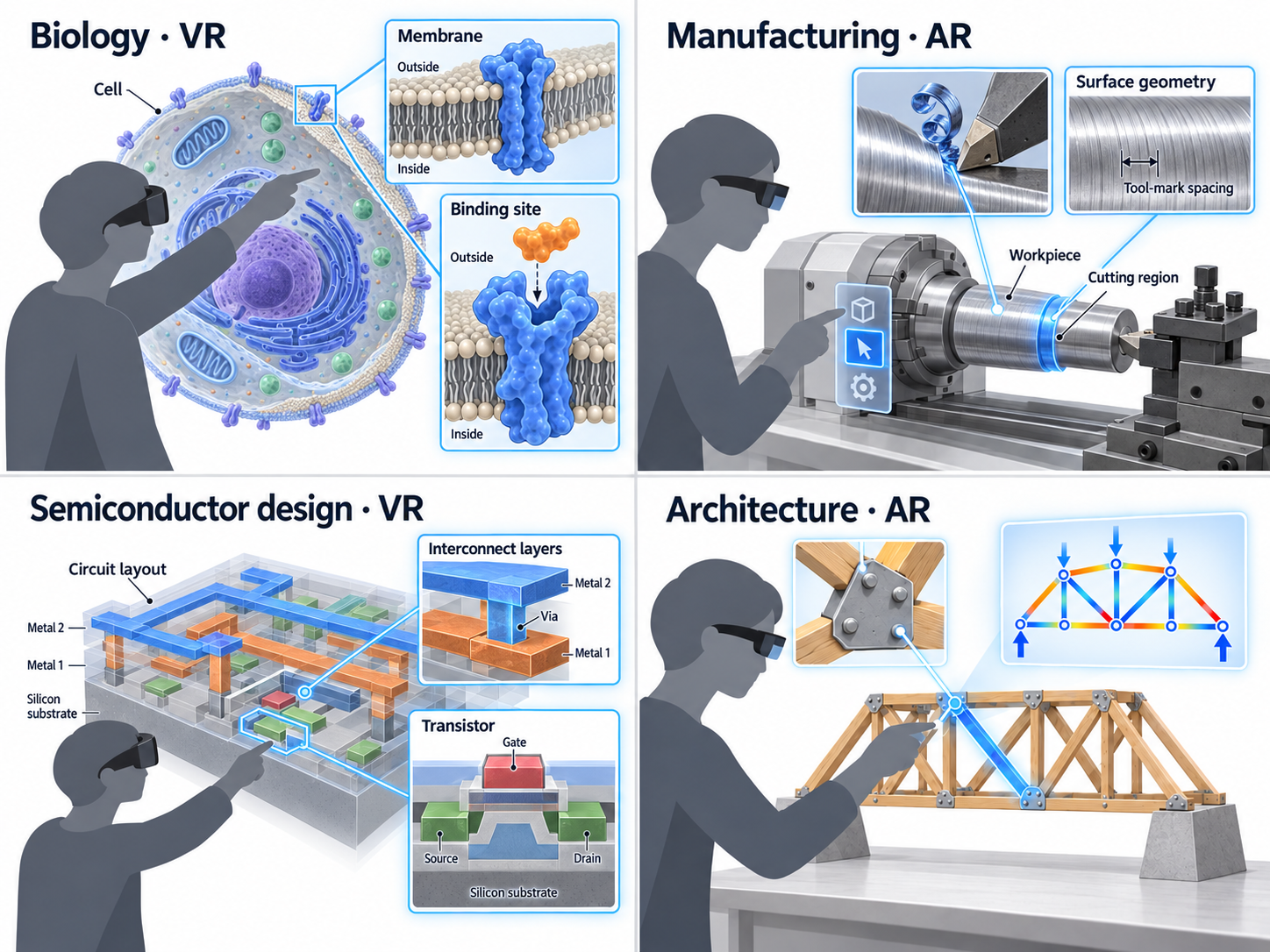}
    \caption{Illustrative applications of Cross-Scale Spatial Reasoning (CSSR) across four immersive learning contexts. Biology connects cellular organization, membrane structure, and molecular interactions; manufacturing connects local tool--material interactions to component-level outcomes; architecture connects structural members and joints to system-level load distribution; and semiconductor design connects device and interconnect structures to circuit behavior. Highlighted regions and linked representations illustrate how learners may investigate relationships across levels. These examples depict proposed instructional scenarios rather than implemented or evaluated systems.}
    \label{fig:cssr-domain-applications}
\end{figure}

\subsection{Biology: From Molecular Structure to Cellular Function}
\label{sec:cssr-biology}

Biological systems involve interactions across molecular, cellular, tissue, and organismal levels. Understanding cellular function often requires learners to explain how molecular structures and interactions contribute to processes occurring at higher organizational levels. Relevant information may be distributed across three-dimensional molecular models, pathway diagrams, and representations of cellular organization. Research on complex biological systems has shown that learners may recognize structures without adequately explaining the behaviors and functions connecting them~\citep{hmelosilver2004sbf,hmelosilver2007fish}.
Existing immersive biology applications demonstrate how interactive three-dimensional environments can make cellular structures accessible for learning. Laubscher and colleagues~\citep{laubscher2024cell}, for example, investigated VR-based learning of cellular structures. Building on such capabilities, CSSR-centered instruction would focus on the relationships connecting molecular structures to cellular behavior rather than the identification of structures alone.
An illustrative activity could investigate how a signaling protein contributes to a cellular response. Learners would examine a molecular model, identify structural features relevant to an interaction, and investigate how the protein participates in a cellular pathway. Linked representations would connect the molecular structure to the larger cellular context, supporting purposeful scale navigation, relational mapping, and representational coordination.
Learners could then investigate a modeled structural modification under clearly specified assumptions. Before observing the simulated outcome, they would predict how the modification might affect molecular interactions and the resulting cellular response. Afterward, they would compare the result with their prediction and explain the relationships and mechanisms responsible. This activity would engage dynamic and causal reasoning alongside consequence-oriented reasoning.
The learning objective would be to explain how molecular features and interactions contribute to cellular function, rather than simply recognize molecular structures or reproduce a pathway sequence. Assessment could introduce an unfamiliar modification and require learners to justify its predicted effects without the correspondence cues or guidance available during instruction. Because biological outcomes may depend on interacting mechanisms and contextual conditions, explanations should also distinguish conclusions supported by the model from those that remain uncertain.

\subsection{Manufacturing: From Local Interactions to Component Quality}
\label{sec:cssr-manufacturing}

Manufacturing processes connect local material interactions to the properties and performance of finished components. In machining, tool geometry, material properties, and operating conditions influence material removal and resulting surface characteristics. Learners may understand the procedural sequence of an operation without explaining how local tool--material interactions contribute to component-level outcomes.
Existing VR-based manufacturing systems provide opportunities for interactive training. Conesa and colleagues~\citep{conesa2024lathe} investigated simulated lathe-operation training, while Ahmed and colleagues~\citep{ahmed2026additive} examined VR-based instruction in additive manufacturing. Such systems establish relevant precedents for immersive manufacturing education, although successful procedural training does not necessarily demonstrate the ability to explain relationships across scales.
A CSSR-centered machining activity could focus on how cutting conditions influence component quality. Learners would examine the overall manufacturing operation, inspect a model of the local tool--material interaction, and connect the relevant process variables to the resulting component characteristics. Navigation between the overall operation, the cutting region, and representations of the finished surface would help learners identify and relate the levels needed to explain the outcome.
Learners could then modify a process parameter, such as feed rate or cutting speed, within a specified range. Before observing the simulation, they would predict how the change might influence local material behavior and component-level properties. They would subsequently compare the modeled outcomes and explain the mechanisms connecting the parameter change to the resulting characteristics.
This approach distinguishes cross-scale understanding from parameter optimization through trial and error. A learner may identify a configuration that produces an acceptable result without understanding the mechanisms responsible. CSSR-centered instruction would instead require learners to connect local interactions to larger-scale outcomes and justify their decisions through the relevant relationships.
Assessment could introduce an unfamiliar material or machining configuration and ask learners to predict its likely effects under specified assumptions. Responses would be evaluated for the accuracy of cross-scale relationships, the adequacy of causal explanations, and the justification of predicted consequences.

\subsection{Cross-Domain Synthesis and Instructional Implications}
\label{sec:cssr-cross-domain-comparison}

The biology and manufacturing examples illustrate how CSSR can guide learning activities involving different scientific mechanisms. In both cases, learners must connect local interactions to broader outcomes, coordinate representations, and use their understanding to explain or predict system behavior. Semiconductor design and architecture provide additional examples of how these reasoning demands can arise in other educational contexts.
In semiconductor education, learners may need to relate device characteristics and physical interconnections to circuit-level performance. Existing immersive systems have supported microfabrication training and visualization of electrical circuit principles~\citep{wang2024cleanroom,wolbach2024circuits}. A CSSR-centered activity could extend these capabilities by requiring learners to identify relationships among device structures, circuit schematics, physical layouts, and timing behavior, as illustrated in Sections~\ref{sec:cssr-worked-example} and~\ref{sec:cssr-design-specification}.
Architecture and structural engineering involve related reasoning demands. Local connections and structural members influence the behavior of larger assemblies, while forces and load paths are not always apparent from geometry alone. AR and VR systems have supported structural analysis and the visualization of forces, deflections, and three-dimensional structural relationships~\citep{turkan2017structural,mojidra2025mola,rodgers2026building}. A CSSR-centered activity could ask learners to trace how a local structural modification affects connected elements, explain the mechanisms governing load distribution, and justify predictions about system-level behavior.
Table~\ref{tbl:cssr-domain-examples} summarizes how the framework can be translated into learning objectives, instructional activities, and assessment evidence across the four domains.

\begin{table*}[t]
\centering
\caption{Illustrative applications of CSSR across four domains, showing the relationships between cross-scale learning objectives, reasoning demands, instructional activities, and assessment evidence.}
\label{tbl:cssr-domain-examples}
\footnotesize
\setlength{\tabcolsep}{5pt}
\renewcommand{\arraystretch}{1.35}
\begin{tabularx}{\textwidth}{@{} >{\raggedright\arraybackslash}p{0.14\textwidth} >{\raggedright\arraybackslash}p{0.21\textwidth} >{\raggedright\arraybackslash}p{0.22\textwidth} >{\raggedright\arraybackslash}X >{\raggedright\arraybackslash}p{0.18\textwidth} @{}}
\toprule
\textbf{Domain} & \textbf{Learning objective} & \textbf{CSSR reasoning demands} & \textbf{Proposed activity} & \textbf{Evidence of understanding} \\
\midrule
\textbf{Biology} & Connect molecular interactions to cellular function. & Relational mapping; representational coordination; causal reasoning. & Investigate signaling through linked molecular and cellular pathway models. & Explain and predict changes in a cellular response. \\
\midrule[0.25pt]
\textbf{Manufacturing} & Connect local process conditions to component properties. & Scale navigation; causal and consequence-oriented reasoning. & Modify machining parameters and investigate modeled material interactions and component outcomes. & Justify the effects of an unfamiliar configuration. \\
\midrule[0.25pt]
\textbf{Semiconductor design} & Connect physical implementation to circuit behavior. & Relational mapping; representational coordination; causal reasoning. & Trace signal paths across schematic, layout, and timing views. & Explain delay and predict the effects of a layout change. \\
\midrule[0.25pt]
\textbf{Architecture} & Connect local structural relationships to system-level behavior. & Scale navigation; relational mapping; consequence-oriented reasoning. & Investigate structural connections and modeled load paths across an assembly. & Explain how local changes influence force distribution. \\
\bottomrule
\end{tabularx}
\end{table*}

These applications demonstrate how a common set of reasoning processes can guide instructional design while accommodating disciplinary differences. Across the four domains, the framework directs attention toward identifying relevant levels, establishing relationships, connecting representations, explaining mechanisms, and predicting the consequences of changes. However, the scientific knowledge needed to perform these tasks is domain-specific, as are the representations and criteria for a correct explanation.
The framework therefore does not imply that cross-scale reasoning develops identically across disciplines or that competence in one domain automatically transfers to another. Its contribution is a common instructional vocabulary for specifying learning objectives, identifying reasoning difficulties, designing support, and assessing explanations across different kinds of complex systems. Whether the proposed instructional mechanisms improve learning, and under what conditions reasoning transfers to unfamiliar problems, remains a question for empirical investigation.

%% file: sections/06-Evaluation.tex
\section{Assessing Cross-Scale Reasoning and Evaluating Instructional Effectiveness}
\label{sec:cssr-assessment-validation}

The value of CSSR as a conceptual and instructional framework depends on two related but distinct questions: whether its proposed reasoning processes can be meaningfully assessed and whether learning activities designed around these processes improve understanding. The first requires evidence that assessment tasks elicit the intended reasoning and that learners' responses can be interpreted reliably and validly. The second requires empirical comparisons that establish whether particular instructional mechanisms support cross-scale understanding relative to appropriate alternatives.
This section outlines an approach to observing and assessing CSSR, examining its relationship to existing cognitive constructs, evaluating the proposed instructional mechanisms, and investigating retention and transfer. The methods described constitute a research agenda rather than an empirically validated assessment instrument or intervention.

\subsection{Observing Cross-Scale Reasoning}
\label{sec:cssr-process-assessment}

Assessing CSSR requires examining how learners establish and use spatial and structural relationships across physical scales or organizational levels. Successful navigation, accurate identification of components, or completion of an interactive task does not necessarily demonstrate cross-scale understanding. Stronger evidence comes from learners' ability to explain why particular levels are relevant, identify relationships connecting them, combine information across representations when necessary, explain mechanisms, and justify predictions about the consequences of changes.
The five CSSR processes provide complementary perspectives for observing this reasoning. Scale navigation can be examined through learners' selection and justification of relevant levels. Cross-scale relational mapping can be assessed through their explanations of structural, functional, or causal connections between levels. Representational coordination can be observed when learners interpret corresponding information across different views. Dynamic and causal reasoning concerns their ability to explain how interactions produce system behavior, while consequence-oriented reasoning concerns their ability to justify predicted outcomes under changed conditions.
These processes can be investigated through written explanations, annotated diagrams, structured interviews, think-aloud protocols, and interactive problem-solving tasks. Different formats may reveal different aspects of reasoning. For example, an annotated diagram may make structural relationships explicit, while a verbal explanation may reveal how a learner interprets the mechanisms connecting those structures. Assessment should therefore select evidence appropriate to the learning objective rather than assume that a single response format captures all aspects of CSSR.

A central methodological challenge is distinguishing incomplete knowledge of individual levels from difficulty integrating information across them. A learner cannot explain a molecular pathway without sufficient biological knowledge, just as a learner cannot accurately explain circuit delay without understanding relevant electrical principles. Assessments must therefore establish whether learners possess the prerequisite knowledge needed to construct the intended explanation.
One approach is to combine within-level and cross-level tasks. Learners could first demonstrate their understanding of individual structures, representations, or domain principles, followed by tasks requiring them to establish relationships across levels. Comparing performance across these tasks may help identify whether difficulties arise from insufficient domain knowledge, unsuccessful integration, or both. Such comparisons would also provide a basis for determining which aspects of reasoning require instructional support.

\subsection{Developing an Assessment Framework}
\label{sec:cssr-assessment-development}

Research on size-and-scale cognition provides useful precedents for assessment development. Magana and colleagues~\citep{magana2012fs2c} proposed a framework for characterizing qualitative and quantitative size-and-scale reasoning, while Harper-Gampp and colleagues~\citep{harpergampp2026assc} developed and validated the Assessment of Size and Scale Cognition (ASSC). These contributions demonstrate the importance of translating theoretical definitions into observable tasks and establishing evidence for the interpretation of assessment outcomes.
CSSR addresses an overlapping but broader set of reasoning demands. Whereas size-and-scale assessments emphasize understanding magnitude, proportion, and relationships among scales, CSSR-centered tasks additionally examine how learners establish and use spatial and structural relationships across organizational levels to explain system behavior and predict consequences. The objective is not to replace existing assessments, but to investigate aspects of cross-scale integration that may not be captured by measures focused on individual spatial or scale-related abilities.
Table~\ref{tbl:cssr-scoring} presents illustrative performance descriptors aligned with the five proposed CSSR processes. The descriptors distinguish incomplete from stronger evidence of cross-scale reasoning, but do not constitute a complete scoring rubric or validated measurement instrument. Intermediate performance levels, scoring rules, and procedures for interpreting results would need to be established through empirical assessment development.

\begin{table*}[t]
\centering
\caption{Illustrative performance descriptors for evaluating the five proposed CSSR reasoning processes. These descriptors are intended to guide assessment development and have not been empirically validated.}
\label{tbl:cssr-scoring}
\footnotesize
\setlength{\tabcolsep}{6pt}
\renewcommand{\arraystretch}{1.35}
\begin{tabularx}{\textwidth}{@{}
>{\raggedright\arraybackslash}p{0.25\textwidth}
>{\raggedright\arraybackslash}X
>{\raggedright\arraybackslash}X @{}}
\toprule
\rowcolor{blue!10}
\textbf{CSSR process} & \textbf{Incomplete evidence} & \textbf{Stronger evidence} \\
\midrule
\textbf{Scale navigation} & Relevant levels are omitted, selected without justification, or treated as interchangeable. & Relevant levels are identified, their roles are explained, and their selection is justified. \\
\midrule[0.25pt]
\textbf{Cross-scale relational mapping} & Structures are described separately or relationships between levels are inaccurate or incomplete. & Accurate structural or functional relationships are established and their significance is explained. \\
\midrule[0.25pt]
\textbf{Representational coordination} & Representations are interpreted independently or their correspondences are misunderstood. & Information from relevant representations is coordinated to support an integrated explanation. \\
\midrule[0.25pt]
\textbf{Dynamic and causal reasoning} & An outcome is described without identifying the mechanisms or interactions responsible. & Relevant mechanisms are connected across levels to explain how system behavior arises. \\
\midrule[0.25pt]
\textbf{Consequence-oriented reasoning} & Predictions lack justification, overlook relevant relationships, or ignore important conditions. & Predictions follow from cross-scale explanations and account for relevant constraints and uncertainty. \\
\bottomrule
\end{tabularx}
\end{table*}

Developing a CSSR assessment should begin with authentic disciplinary problems in which successful performance requires reasoning across levels. Domain experts can evaluate the scientific accuracy and relevance of the tasks, while interviews and think-aloud studies can help reveal how learners interpret the problems and construct their responses. Scoring criteria should then be refined through analysis of learner responses, with attention to inter-rater consistency, task difficulty, and the extent to which the criteria distinguish meaningful differences in reasoning.
The relationships among the five processes must also be investigated rather than assumed. In authentic learning tasks, accurate causal explanations may depend on successful relational mapping, while representational coordination may contribute to several other processes. Assessment research should therefore examine whether the processes can be meaningfully distinguished in learner performance, whether they are best represented as related dimensions, and whether their relationships differ across tasks and domains.

\subsection{Establishing Construct Validity}
\label{sec:cssr-construct-validity}

Establishing the validity of CSSR assessment requires evidence supporting the intended interpretations and uses of assessment results. Reliability is necessary for consistent measurement, but reliable scores alone do not establish that an assessment captures the proposed reasoning. Established approaches to construct validity emphasize the alignment among theoretical definitions, task content, response processes, assessment structure, and relationships with other measures~\citep{cronbach1955validity,messick1994validity,aera2014standards}.
This distinction is particularly important because CSSR integrates processes already examined in spatial cognition, scale reasoning, representational competence, and systems thinking. Initial validation studies could investigate how performance on CSSR-centered tasks relates to established measures of these abilities and to learners' disciplinary knowledge. Such comparisons would help determine which aspects of cross-scale performance are explained by existing constructs and whether the integrated framework provides additional explanatory value.
For example, researchers could examine whether learners' ability to construct cross-scale explanations predicts performance on unfamiliar problems after accounting for prior domain knowledge and spatial reasoning ability. If CSSR-centered assessments provide additional information about learners' explanations or predictions, this would support the framework's explanatory usefulness. If performance is largely accounted for by existing measures, CSSR may be better understood as an instructional framework that coordinates established abilities rather than as a psychometrically distinct cognitive construct.

Importantly, the value of CSSR does not depend exclusively on demonstrating that it represents a new independent ability. The framework may also contribute by identifying recurring patterns of reasoning difficulty, organizing instructional objectives, and supporting the design of targeted interventions. Construct-validation research should therefore examine both the proposed organization of reasoning processes and the practical explanatory value of that organization.
Validation must also consider disciplinary context. Tasks involving cellular signaling, machining, circuit design, or structural analysis may engage similar reasoning processes but require different scientific knowledge and representations. Shared assessment dimensions may be possible, yet equivalent interpretations of scores across domains cannot be assumed. Evidence from multiple disciplinary contexts would be needed to determine which aspects of CSSR generalize and which remain dependent on domain-specific expertise.

\subsection{Evaluating Instructional Mechanisms}
\label{sec:cssr-intervention-evaluation}

The research propositions introduced in Section~\ref{sec:cssr-propositions} provide a basis for evaluating the instructional implications of CSSR. These propositions concern the effects of relational continuity, coordinated representations, prediction and explanation, targeted scaffolding, and explicit cross-scale instruction. Empirical studies should distinguish the contributions of these mechanisms from the effects of immersive presentation itself.

To examine relational continuity (P1), researchers could compare transitions between levels that preserve relevant spatial or structural relationships with transitions that provide equivalent information without explicit relational cues. Learners' explanations and reconstructions of cross-scale relationships could be evaluated to determine whether preserving context improves integration, particularly among learners who understand individual levels but struggle to connect them.
Coordinated representations (P2) could be investigated by comparing linked views that communicate meaningful correspondences with independently presented representations containing the same scientific information. The primary outcome would be learners' ability to establish accurate relationships and integrate complementary information. Such studies should account for prior representational knowledge, because learners must understand individual views before they can effectively coordinate them.
Prediction and explanation activities (P3) could be evaluated by comparing instruction that requires learners to predict an outcome, investigate it through simulation, and explain the mechanisms involved with observation-focused activities using the same content and simulation. Assessment should emphasize the quality of causal explanations and justified predictions, including performance on unfamiliar problems, rather than success on the practiced simulation alone.
Targeted scaffolding (P4) could be studied by comparing guidance matched to an identified reasoning difficulty with general guidance of comparable detail. For example, learners struggling with relational mapping might receive correspondence cues, while those struggling with causal explanation might receive mechanism-focused prompts. Researchers should examine whether the targeted guidance improves the corresponding aspect of reasoning and whether those benefits persist as support is gradually removed.
Transfer (P5) could be examined by comparing instruction that explicitly emphasizes relationships connecting levels with instruction focused on individual structures or representations. Learners could subsequently complete unfamiliar problems that preserve relevant relational principles while changing the surface features, representations, or configurations. These comparisons would help determine whether explicitly developing cross-scale relationships supports the application of understanding beyond practiced examples.

Across these studies, researchers should account for prior knowledge, instructional time, task difficulty, and the scientific information available across conditions. Written explanations, annotated diagrams, interviews, and interactive problem-solving tasks can provide direct evidence of reasoning, while interaction records, cognitive load, usability, and engagement measures can offer complementary information. Comparisons between immersive and non-immersive implementations may further establish whether particular advantages arise from AR or VR affordances or from instructional mechanisms that could also be implemented through other media.

\subsection{Retention and Transfer}
\label{sec:cssr-transfer}

An important educational question is whether learners retain cross-scale understanding and apply it beyond the particular examples and interactions encountered during instruction. Immediate success may reflect familiarity with a representation, simulation, or practiced sequence rather than the ability to establish relationships independently. Retention and transfer should therefore be examined as distinct outcomes.
Retention concerns whether learners can reconstruct relationships, explain mechanisms, and justify predictions after a delay and without the guidance available during instruction. Delayed assessments could present previously studied phenomena through altered representations or require learners to produce explanations independently. Such tasks would help determine whether instructional benefits extend beyond immediate performance.

Within-domain transfer concerns applying cross-scale reasoning to unfamiliar problems within the same disciplinary context. For example, a learner who has investigated one circuit configuration might explain a different timing problem, while a manufacturing learner could predict the consequences of a new material or machining configuration. These tasks should preserve relevant explanatory principles while varying features that prevent learners from reproducing a memorized solution.

Cross-domain transfer presents a more demanding question. Although CSSR proposes reasoning processes that recur across scientific and engineering disciplines, applying them in a different domain requires knowledge of that domain's structures, mechanisms, and representational conventions. Research on transfer emphasizes the importance of specifying which task features and underlying relationships remain consistent between learning and assessment contexts~\citep{barnett2002transfer,goldstone2008transfer}. Studies should therefore distinguish the transfer of general reasoning strategies from the application of domain-specific scientific knowledge and avoid interpreting unsuccessful cross-domain performance as evidence of a general reasoning deficit without considering prerequisite knowledge.

Together, evidence concerning assessment validity, instructional effectiveness, retention, and transfer would help establish the educational value and limitations of CSSR. The central research objective is not simply to determine whether immersive learning environments improve performance, but to identify which forms of instructional support help learners establish, explain, and apply relationships across scales and organizational levels, for whom, and under what conditions.

%% file: sections/07-Discussion.tex
\section{Discussion}
\label{sec:discussion}

The proposed CSSR framework brings together established perspectives on spatial cognition, scale reasoning, multiple representations, and systems thinking to address a recurring challenge in learning about complex systems: learners may understand structures and processes at individual levels without being able to connect them into an explanation of system behavior. By characterizing the reasoning involved in establishing these connections, CSSR provides a basis for identifying learning difficulties and aligning instructional design with the understanding learners are expected to develop. Its implications extend beyond how complex systems are visualized to how cross-scale relationships can be taught, examined, and assessed.

\subsection{Theoretical Implications of Cross-Scale Integration}
\label{sec:discussion-theory}

A central implication of CSSR is that difficulties in understanding complex systems may arise from different forms of incomplete integration, even when learners possess relevant knowledge of individual components or representations. A learner may recognize structures at several levels but fail to identify their relationships, coordinate different representations without explaining the mechanisms connecting them, or understand an observed outcome without being able to predict the consequences of a change. Although these difficulties may produce similarly incomplete explanations, they involve different reasoning demands and may require different forms of instructional support.
 
Existing research provides important foundations for understanding these demands. Studies of thinking in levels examine relationships between local interactions and larger-scale behavior~\citep{wilensky1999levels,jacobson2006complex}. Research on size-and-scale cognition addresses magnitude and proportional relationships~\citep{magana2012fs2c,cheek2017scale}, while multiple-representation frameworks examine how learners interpret and coordinate complementary information~\citep{ainsworth2006deft,rau2017representations}. CSSR builds on these perspectives by organizing their contributions around tasks in which spatial and structural relationships across levels must be used to construct explanations and predictions.
This integrated account has implications for how cross-scale learning difficulties are interpreted. Rather than treating unsuccessful performance as a general failure of spatial understanding or domain knowledge, the framework encourages closer examination of the relationships learners can identify, the information they can coordinate, and the mechanisms they can explain. Such distinctions may help researchers identify patterns of reasoning that remain difficult even when learners demonstrate adequate understanding of individual levels.
 
The theoretical contribution of CSSR therefore lies in characterizing a class of reasoning demands and their coordination, rather than claiming the discovery of entirely new cognitive operations. Whether these processes form a distinguishable cognitive construct remains an empirical question. Even if future research finds substantial overlap with existing measures, the framework may retain value as an account of how established abilities contribute to cross-scale learning tasks and how their coordination can inform instruction.

\subsection{Implications for Immersive Learning Design}
\label{sec:discussion-design}

The framework suggests a shift in how immersive learning environments are conceptualized and evaluated. AR and VR offer capabilities for navigating spatial structures, coordinating representations, manipulating objects, and visualizing otherwise inaccessible processes. However, these capabilities do not determine which relationships learners understand or whether they can use those relationships to explain system behavior. CSSR places the reasoning objective, rather than the immersive interaction, at the center of instructional design.
This distinction is particularly important for learning tasks that involve multiple scales or organizational levels. An immersive environment may allow learners to move seamlessly from a system-level view to a detailed representation of its components, yet the transition may provide little instructional benefit if the relationships between those levels are unclear. Similarly, synchronized views may improve access to complementary information without necessarily helping learners construct causal explanations. The instructional value of these mechanisms depends on whether they support the specific reasoning demands of the task.
CSSR therefore encourages designers to consider not only what learners can see or manipulate, but also what they must understand and explain. Navigation mechanisms can preserve relationships across levels, coordinated representations can make correspondences explicit, simulations can support investigation of causal mechanisms, and contextual feedback can help learners reconsider incomplete explanations. Importantly, these mechanisms may support several reasoning processes simultaneously, and their effectiveness is likely to depend on learners' prior knowledge and the complexity of the system being studied.
 
The framework also has implications for adaptive learning environments. Interaction traces such as navigation patterns, time spent examining structures, or the frequency of manipulating representations may reveal aspects of learner activity, but do not necessarily establish what learners understand. Adaptation informed by CSSR would require stronger evidence of reasoning, including the relationships learners identify, the explanations they construct, and the predictions they can justify. Such evidence could help distinguish whether a learner needs support in selecting relevant levels, coordinating representations, or explaining mechanisms, rather than treating all unsuccessful interactions as equivalent.
A further implication is that CSSR does not assume immersive technology is always the most appropriate instructional medium. Cross-scale understanding may also be supported through linked diagrams, physical models, or interactive desktop simulations when these provide sufficient access to the relevant relationships. The value of immersion should therefore be examined in relation to the learning demands and the specific affordances it offers, rather than treated as inherently beneficial. In this respect, CSSR complements models such as CAMIL, which examine how immersive characteristics influence learning through cognitive and affective processes~\citep{makransky2021camil}. CSSR focuses on the cross-scale reasoning that instructional activities are intended to develop.

\subsection{Scope and Limitations}
\label{sec:discussion-limitations}

CSSR is proposed as a conceptual and instructional framework and has not yet been empirically validated. Its five constituent processes were identified through theoretical synthesis, and their relationships, measurement properties, and degree of independence remain to be established. The instructional principles and research propositions similarly require empirical evaluation before conclusions can be drawn about their effectiveness.
The framework is also bounded by the kinds of learning tasks it seeks to explain. Not all spatial reasoning involves relationships across scales, and not all complex-system reasoning depends on spatial or structural information. Furthermore, changes in physical scale, organizational level, and representation are not interchangeable, even when they occur together within a learning activity. The framework must therefore be applied with attention to the particular relationships and explanatory demands of a task.
 
Another limitation concerns the role of disciplinary knowledge. Biology, manufacturing, semiconductor design, and architecture involve different scientific mechanisms, representational conventions, and standards for accurate explanations. Similar reasoning demands across these domains do not establish that learners possess a domain-general reasoning ability or that proficiency will transfer readily between disciplines. The illustrative applications presented in this paper are proposed instructional scenarios rather than implemented CSSR interventions. Although existing immersive learning research provides relevant technological and pedagogical precedents, its findings should not be interpreted as validation of the proposed framework.

\subsection{Future Research Directions}
\label{sec:discussion-future}

An immediate research priority is to examine how the five proposed CSSR processes manifest in authentic learning tasks. Studies of learners' explanations, representational choices, and problem-solving strategies could investigate whether the processes describe recurring patterns of reasoning difficulty and how these patterns relate to existing measures of spatial cognition, scale understanding, and systems thinking. Such work would help establish whether CSSR provides additional explanatory value and inform the development of valid assessment approaches.
 
A second direction concerns the relationship between reasoning demands and instructional support. The framework proposes that different difficulties may benefit from different mechanisms, including relational cues, coordinated representations, prediction and explanation activities, and targeted scaffolding. Future research should determine when these mechanisms improve cross-scale understanding, how their effectiveness depends on prior knowledge, and whether their benefits remain when support is removed. In particular, comparisons across immersive and non-immersive implementations could clarify when the spatial and interactive capabilities of AR and VR offer educational advantages.
 
A longer-term direction is to investigate how cross-scale reasoning develops through learning and experience. Longitudinal studies could examine whether learners become more proficient at selecting relevant levels, establishing relationships, and constructing explanations as they gain disciplinary knowledge. Research on transfer could further investigate whether learners apply these processes to unfamiliar problems within a domain and under what conditions similar reasoning strategies can be used across disciplines.
The framework also raises questions about the design of adaptive immersive learning environments. If cross-scale reasoning difficulties can be identified through reliable assessment, instructional systems may be able to provide support responsive to learners' explanations and predictions. Developing such systems would require methods that distinguish meaningful evidence of understanding from interaction behavior alone, as well as investigations of when adaptive guidance supports learning without reducing learners' opportunities for independent reasoning.
Together, these directions position CSSR as a research framework whose theoretical and instructional value can be progressively examined. Its longer-term contribution will depend on whether it helps researchers identify meaningful patterns of cross-scale reasoning, develop more effective instructional support, and establish what learners can explain and apply beyond the representations and activities used during instruction.

%% file: sections/08-conclusion.tex
\section{Conclusion}
\label{sec:conclusion}

The challenge of learning complex scientific and engineering
systems lies not only in understanding their individual
components, but also in recognizing how relationships
across scales shape the behavior of the system as a whole.
Immersive technologies create opportunities to explore
these relationships, yet making structures visible and
interactive is only a starting point. Meaningful learning
requires learners to connect what they observe across
levels, explain the mechanisms underlying system
behavior, and apply that understanding to unfamiliar
problems.
Cross-Scale Spatial Reasoning (CSSR) offers a framework
for making these learning demands explicit. By connecting
cognitive processes with instructional design and
assessment, it provides a basis for developing immersive
experiences that emphasize reasoning rather than
visualization alone. Future research must establish how
these processes develop, which forms of instructional
support are most effective, and whether the resulting
understanding transfers across problems and domains.
Ultimately, the promise of immersive learning lies not
simply in allowing learners to explore complex systems
at different scales, but in helping them understand
the relationships that connect those scales and
give rise to system behavior.